\documentclass[11pt,twoside]{article}
\usepackage{asp2006}
\usepackage{epsf}
\usepackage{lscape}

\begin{document}
\title{On determining the mass-loss rates of red giants and red supergiants
based on infrared data}
\author{Jacco Th. van Loon}
\affil{School of Physical \& Geographical Sciences,
       Lennard Jones Laboratories,
       Keele University,
       Staffordshire ST5 5BG,
       UK
       (jacco@astro.keele.ac.uk)}

\begin{abstract}
I review existing methods for determining mass-loss rates of red giants and
red supergiants based on infrared data. The simplest method is based on models
for the absorption and emission by dust which forms in the dense outflows from
these cool stars. I discuss the parameters and assumptions upon which the
method relies, review relationships between the mass-loss rate and infrared
colours or far-infrared flux density, and propose a new formula for the
mass-loss rate as a function of the visual extinction. I also briefly discuss
the use of atomic and molecular transitions at infrared wavelengths.
\end{abstract}

\section{Introduction}

Early spectroscopic observations of cool giant stars revealed violet-displaced
absorption in the optical line profiles of strong electronic transitions of
singly-ionized and neutral atoms, and mass-loss rates of $\dot{M}\sim10^{-8}$
to $10^{-6}$ M$_\odot$ yr$^{-1}$ were inferred
\citep{Deutsch1956,Reimers1975}. More evidence of such winds came in the form
of radio emission lines from abundant molecules, in particular the amplified
stimulated emission from hydroxyl at $\lambda=18$ cm \citep*{ElitzurEtal1976}.
Thermal line emission from carbon monoxide at mm wavelengths enabled
measurements of the wind speed and mass-loss rate for carbon stars as well as
stars with oxygen-rich envelopes \citep{KnappMorris1985}. Methods for
determining the mass-loss rate using radio techniques are described by Fredrik
Sch\"oier elsewhere in these proceedings.

Infrared surveys revealed stars which are heavily reddened by circumstellar
dust. The association between these dust-enshrouded stars and masers was
quickly made, and the accompanying dust emission became a very useful tracer
of mass loss \citep{GehrzWoolf1971}. It also explained the pumping of the
masers through infrared emission, but most importantly it provided a mechanism
for driving the wind. Strong pulsation of the cool photospheres of these stars
acts as a piston to increase the scaleheight of the molecular atmosphere,
facilitating the condensation of grains \citep{Jura1986}. The continuum
opacity of the grains allows for efficient transfer of momentum from the
stellar radiation field onto the dust grains. If the density is high enough
then the dust and gas are coupled, mostly via grain-H$_2$ collisions. High
mass-loss rates were estimated for these stars, $\dot{M}\sim10^{-6}$ to
$10^{-4}$ M$_\odot$ yr$^{-1}$ \citep{OlnonEtal1984}.

The {\it Spitzer Space Telescope} has made it possible to detect the infrared
emission from circumstellar dust around red giants and red supergiants (RSGs)
in many Local Group galaxies for the first time. This enables the study of
mass loss and dust production in a rich variety of environments inaccessible
before, for example in the Wolf-Lundmark-Melotte dwarf galaxy
\citep{JacksonEtal2007} but also galactic globular clusters such as the
extremely metal-poor Messier 15 \citep{BoyerEtal2006}, whilst more thorough
and detailed studies are possible in the Magellanic Clouds. As technological
progress inflates the observable Universe, much of the groundbraking work is
done in uncharted territory and at the limit of sensitivity, running into many
of the types of problems that were faced in the 1970s in studies of nearby
stars. It is thus essential to understand the methods that are applied to
measure mass-loss rates from these infrared data, and in particular to be
aware of their assumptions and limitations.

\section{Dust-driven wind model as the basis for measuring mass-loss rates}

Circumstellar dust grains absorb stellar light mainly at optical (and
ultraviolet) wavelengths, and re-emit it mainly at infrared wavelengths. The
shape of the observed spectral energy distribution (SED) depends on the
optical properties of the grains and on the optical depth of the envelope. For
a spherically symmetric geometry the integral under the SED yields the
bolometric luminosity, provided that the distance is known. The SED does {\it
not} allow a direct measurement to be made of the mass-loss rate. Under the
assumption of radiative equilibrium between the capture of photons and
isotropic emission by the heated grain, and applying the continuity equation,
one obtains a crude relationship that summarises the problem quite well
\citep{IvezicElitzur1995}:
\begin{equation}
\tau \propto \frac{\psi \dot{M}}{v_{\rm exp} \sqrt{L}},
\end{equation}
where we notice that although the optical depth, $\tau$, is proportional to
the (gas+dust) mass-loss rate, $\dot{M}$, it also depends on the dust:gas mass
ratio, $\psi$, the expansion velocity of the wind, $v_{\rm exp}$, as well as
the luminosity, $L$.

Simple radiation-driven dust wind theory predicts how the wind speed should
depend on the luminosity and dust:gas ratio. This can be used to compute the
expected wind speed for stars for which we have no direct measurement of it.
The momentum equation relates the motion of the matter and photon fluids:
\begin{equation}
\dot{M} v_{\rm exp} \propto \tau L,
\end{equation}
where the optical depth properly accounts for the scattering of photons off
the circumstellar grains. Hence, combination with Eq.\ (1) yields:
\begin{equation}
v_{\rm exp} \propto \sqrt{\psi} \sqrt[4]{L}.
\end{equation}
It seems deceptively reasonable that the dust:gas ratio depends on
metallicity. Unfortunately, this is a difficult parameter to measure directly.

First evidence for the metallicity dependence of the wind speed of red giants
was presented by \citet{WoodEtal1992} who detected six OH/IR stars in the LMC.
\citet{MarshallEtal2004} enlarged this sample, and from a comparison with the
wind speeds and luminosities of OH/IR stars in the galactic centre they could
confirm Eq.\ (3) and indicate that the dust:gas ratio is linearly proportional
to metallicity:
\begin{equation}
\psi \propto Z.
\end{equation}
In the absence of direct measurements of wind speed and dust:gas ratio, it is
therefore recommended to abide by the following scaling relation:
\begin{equation}
\frac{\dot{M}}{\tau} \propto Z^{-0.5} L^{0.75},
\end{equation}
The metallicity dependence of dust-driven winds is reviewed in
\citet{Vanloon2006}.

\section{Model computations of the infrared spectral energy distribution}

The optical depth of the dust envelope can be derived from the observed SED by
comparison with synthetic SEDs. The latter are produced by computing
simultaneously the radiation transfer through the dust envelope, and the
thermal balance between the grain heating through irradiation and cooling
through reradiation. The principles underlying these computations were first
outlined for the non-grey case by \citet{Leung1975}, \citet{Rowanrobinson1980}
and \citet{Yorke1980}. Three commonly used codes that solve the equations of
radiation transfer and thermal equilibrium were developed independently in the
1990s: the code of \citet{Groenewegen1993}, the {\sc dusty} code
\citep*{IvezicEtal1999}, and the {\sc modust} code \citep{Bouwman2001}. Of
these, only {\sc dusty} is publicly available --- fortunately it is also well
documented and relatively easy to use. The latest generation of codes are
based on Monte Carlo techniques for tracing the energy packets on their
journey through the dust envelope \citep{BjorkmanWood2001}, culminating in the
publicly available photo-ionization/dust radiation transfer code {\sc
mocassin} \citep*{ErcolanoEtal2005}.

\begin{figure}[!t]
\plotone{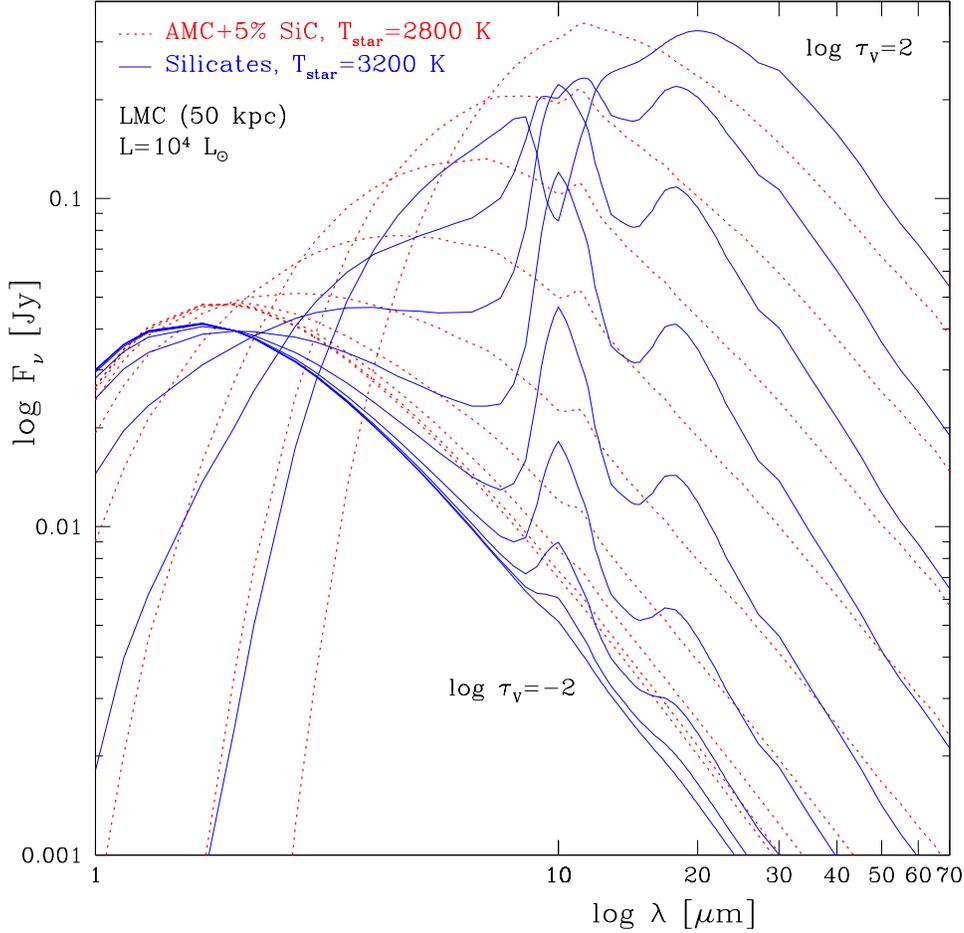}
\caption{Spectral energy distributions computed with the radiation transfer
code {\sc dusty} \citep*{IvezicEtal1999}, for a spherically symmetric envelope
surrounding a blackbody with a luminosity of $10^4$ L$_\odot$ placed at the
distance of the Large Magellanic Cloud. Two cases are considered:
$T_\star=3200$ K and oxygenous grains (silicates), and $T_\star=2800$ K and
carbonaceous grains (amorphous carbon and 5\% admixture of silicon-carbide).
The computations are performed on a logarithmic grid of optical depth from
$\tau_{\rm V}=0.01$ to 100.}
\end{figure}

As an illustration, the {\sc dusty} code is used to generate a series of model
SEDs with a visual optical depth between $\tau_{\rm V}=0.01$ and 100, for an
oxygen-rich red giant and a carbon star (Fig.\ 1). Corresponding mass-loss
rates are $\dot{M}\sim5\times10^{-8}$ to $4\times10^{-5}$ M$_\odot$ yr$^{-1}$
for a typical Asymptotic Giant Branch (AGB) star at solar metallicity, but
note that the difference in luminosity between massive RSGs and the tip of the
Red Giant Branch corresponds to a two orders of magnitude difference in the
mass-loss rate inferred from the shape of the SED alone.

Differences in optical depth manifest themselves most through extinction of
stellar light at $\lambda<3$ $\mu$m, the strength of discrete features
associated with certain minerals mainly in the $\lambda=10$ to 30 $\mu$m
region, and a general elevation of the spectrum over that of a naked star at
$\lambda>30$ $\mu$m. The extinction only becomes noticeable when $\tau_{\rm
V}$ approaches unity, as it is difficult to isolate its effect from the
depression of the stellar continuum due to blanketing by molecular bands in
cool giants, and from the interstellar contribution to the extinction. Because
the absorbed photons are more energetic than the reradiated photons, small
optical depths readily give rise to detectable amounts of excess emission at
longer wavelengths. In particular, silicates shine efficiently around
$\lambda=10$ $\mu$m, and the resulting emission feature can be used to detect
mass-loss rates as low as a few $10^{-8}$ M$_\odot$ yr$^{-1}$ in
low-luminosity or metal-rich red giants --- but here, too, the sensitivity is
limited by molecular absorption bands and uncertainty in the stellar continuum
(stars are not blackbodies) as well as the detailed mineralogy.

\section{A formula for the mass-loss rate as a function of optical extinction}

For the examples shown in Fig.\ 1, the mass-loss rate turns out to be related
to the visual extinction, $A_{\rm V}$ ($=1.086\,\tau_{\rm V}$), as
$\dot{M}\propto A_{\rm V}^{0.75}$. Incidentally, replacing $\tau$ by $A_{\rm
V}^{0.75}$ in Eq.\ (5) and calibrating against the computed models yields a
practically identical constant of proportionality for both silicates and
amorphous carbon grains. Hence:
\begin{equation}
\dot{M}[{\rm M}_\odot {\rm yr}^{-1}] = 1.5\times10^{-9}\, Z[{\rm
Z}_\odot]^{-0.5} L[{\rm L}_\odot]^{0.75} A_{\rm V}^{0.75}.
\end{equation}
This provides a new, simple and versatile recipe for estimating mass-loss
rates.

\section{Relationships between mass-loss rate and infrared flux or colour}

As a shortcut to comprehensive modeling of the SED, various relationships have
been suggested between the optical depth and a monochromatic infrared flux or
a single infrared colour. The main reason why this is interesting, is because
it may not be possible or economic to sample all of the SED.

The {\it InfraRed Astronomical Satellite} ({\it IRAS}) made it possible to
measure the long-wavelength tail of the dust emission from galactic red
giants. This was first quantified by \citet{Jura1987}, who proposed a formula
for the mass-loss rate as a function of the flux density at 60 $\mu$m --- it
depends also on the wind speed, distance, luminosity, dust:gas ratio, and the
``mean wavelength of the light emerging from the star and its circumstellar
dust shell'' (which requires measurement of a significant portion of the SED).
The formula was adapted to be used at 25 $\mu$m by \citet*{WhitelockEtal1991}
and \citet{WhitelockEtal1994}. Both formulae compare favourably with
measurements of the mass-loss rate from the CO(J=1$\rightarrow$0) rotational
transition at $\lambda=3$ mm.

Relationships between mass-loss rate and colours are appealing as they are
distance invariant, but in the light of the previous discussion this seems a
rather deceptive advantage as they must depend on the star's luminosity. Such
(and other unaccounted for) dependencies are not fully appreciated when they
are derived for relatively uniform samples of stars, for instance carbon stars
or nearby low-mass red giants in a narrow range of metallicity. Formulae are
considered successful if they yield mass-loss rates to an intrinsic accuracy
of a factor two or so, but the same accuracy cannot be guaranteed if applied
outside its validity domain (not limited by just colour range).

I nonetheless summarise in Table 1 several formulae for the mass-loss rate as
a function of infrared colour. Where a second reference is given the data
presented by the former were first parameterised by the latter. Linear,
square-root or asymptotic functions have been suggested. They share many of
the basic assumptions, and notably they are all evaluated for stars of a
(near) solar metallicity --- even the mass-loss rates for the presumably
metal-poor magellanic stars were derived adopting values for the dust:gas
ratio and wind speed typical for red giants encountered in the solar
neighbourhood.

The approaches of \citet{LesidanerLebertre1996}, \citet{Lebertre1997},
\citet{LebertreWinters1998}, and of \citet{SchutteTielens1989}
\citep[parameterised by][]{ZijlstraEtal1996} and \citet{GroenewegenEtal2007}
(parameterised in this review) are based on modeling of the SEDs to derive
mass-loss rates and then correlating with the near-infrared colours of the
same stars. The approach of \citet{WhitelockEtal1994} \citep[parameterised
by][]{ZijlstraEtal1996} is similar as they use the \citet{Jura1987} formalism
to derive the mass-loss rate, which is a crude way of modeling the SED. They
found that the correlation between the mass-loss rate and 25:60 $\mu$m flux
ratio is rather poor. The mass-loss rates from the infrared data in
\citet{LesidanerLebertre1996}, \citet{Lebertre1997},
\citet{LebertreWinters1998} and \citet{WhitelockEtal1994} were all found to be
in good agreement with the mass-loss rates from the CO(J=1$\rightarrow$0) line
emission for the same stars. The approach of \citet{GuandaliniEtal2006}
(parameterised in this review) differs in that the mass-loss rates that were
correlated with the infrared colours were not in themselves derived from
infrared data but from the CO line emission.

\begin{landscape}
\begin{table}
\caption{Formulae for the mass-loss rate as a function of infrared colour,
with a synopsis of the more pertinent restrictions. Dependencies on luminosity
$L$ [L$_\odot$] are implicit unless noted otherwise. References:
1=\citet{GroenewegenEtal2007},
2=\citet{GuandaliniEtal2006},
3=\citet{Lebertre1997},
4=\citet{LebertreWinters1998},
5=\citet{LesidanerLebertre1996},
6=\citet*{SchroederEtal2003},
7=\citet{SchutteTielens1989},
8=\citet{WhitelockEtal1994},
9=\citet{ZijlstraEtal1996},
10=this review.}
\smallskip
\begin{center}
{\small
\begin{tabular}{rlrrrrllllll}
\tableline
\noalign{\smallskip}
\#                  &
\multicolumn{5}{c}{$\log\dot{M}(x)$ [M$_\odot$ yr$^{-1}$] $=a(x+b)^c+d$} &
validity domain     &
chemistry           &
$\psi$              &
$v_{\rm exp}$       &
remarks             &
ref                 \\
\noalign{\smallskip}
\cline{2-6}
\noalign{\smallskip}
                    &
$x$                 &
$a$                 &
$b$                 &
$c$                 &
$d$                 &
                    &
                    &
                    &
[km s$^{-1}$]       &
                    &
                    \\
\noalign{\smallskip}
\tableline
\noalign{\smallskip}
1                   &
$J$--$K$            &
$-6$                &
$-0.2$              &
$-1$                &
$2$                 &
$2.0<x<7$           &
carbon              &
0.01                &
data,15             &
local Miras         &
3                   \\
\noalign{\smallskip}
2                   &
$J$--$K$            &
$-2.5$              &
$-0.65$             &
$-1$                &
$-4.25$             &
$1.6<x<6$           &
oxygen              &
0.01                &
data                &
local Miras         &
4                   \\
3                   &
$J$--$K$            &
$0.156$             &
$7.4\log L$         &
$1$                 &
$-10.38$            &
$-5.3<\log\dot{M}<-4.66$ &
carbon              &
model               &
model               &
synthetic           &
6                   \\
\noalign{\smallskip}
4                   &
$J$--$K$            &
$0.108$             &
$7.4\log L$         &
$1$                 &
$-8.62$             &
$\log\dot{M}>-4.66$ &
carbon              &
model               &
model               &
synthetic           &
6                   \\
\noalign{\smallskip}
5                   &
$K$--$L^\prime$     &
$-9$                &
$1.4$               &
$-1$                &
$2.75$              &
$1.0<x<8$           &
carbon              &
0.01                &
data,15             &
local Miras         &
3                   \\
\noalign{\smallskip}
6                   &
$K$--$L^\prime$     &
$-2.75$             &
$0$                 &
$-1$                &
$-3.75$             &
$0.7<x<3$           &
oxygen              &
0.01                &
data                &
local Miras         &
4                   \\
\noalign{\smallskip}
7                   &
$K$--$L$            &
$0.356$             &
$3.1\log L$         &
$1$                 &
$-10.45$            &
$-5.3<\log\dot{M}<-4.66$ &
carbon              &
model               &
model               &
synthetic           &
6                   \\
\noalign{\smallskip}
8                   &
$K$--$L$            &
$0.236$             &
$3.1\log L$         &
$1$                 &
$-8.50$             &
$\log\dot{M}>-4.66$ &
carbon              &
model               &
model               &
synthetic           &
6                   \\
\noalign{\smallskip}
9                   &
$\log(K$--$[12])$   &
$5.71$              &
$0$                 &
$1$                 &
$-9.41$             &
$0.3<x<1$           &
oxygen              &
0.01                &
data                &
AGB+RSGs            &
5                   \\
\noalign{\smallskip}
10                  &
$K$--$[12]$         &
$-24$               &
$4$                 &
$-1$                &
$-3$                &
$2<x<14$            &
carbon              &
0.01                &
data                &
local Miras         &
4                   \\
\noalign{\smallskip}
11                  &
$K$--$[12]$         &
$-55$               &
$5$                 &
$-1$                &
$0$                 &
$2<x<7$             &
oxygen              &
0.01                &
data                &
local Miras         &
4                   \\
\noalign{\smallskip}
12                  &
$K$--$[12]$         &
$0.190$             &
$5.85\log L$        &
$1$                 &
$-10.34$            &
$-5.3<\log\dot{M}<-4.66$ &
carbon              &
model               &
model               &
synthetic           &
6                   \\
\noalign{\smallskip}
13                  &
$K$--$[12]$         &
$0.130$             &
$5.85\log L$        &
$1$                 &
$-8.54$             &
$\log\dot{M}>-4.66$ &
carbon              &
model               &
model               &
synthetic           &
6                   \\
\noalign{\smallskip}
14                  &
$K$--$[12]$         &
$0.57$              &
$0$                 &
$1$                 &
$-8.05$             &
$x<5.5$             &
oxygen              &
0.005               &
10                  &
local Miras         &
8,9                 \\
\noalign{\smallskip}
15                  &
$L^\prime$--$[12]$  &
$0.58$              &
$0$                 &
$1$                 &
$-6.57$             &
$2.2<x<4.6$         &
oxygen              &
0.004               &
10                  &
$\log L=4$          &
7,9                 \\
\noalign{\smallskip}
16                  &
$[6.4]$--$[9.3]$    &
$2.55$              &
$0$                 &
$0.5$               &
$-7.6$              &
$x<1.4$             &
carbon              &
0.005               &
10                  &
magellanic          &
1,10                \\
\noalign{\smallskip}
17                  &
$K_{\rm s}$--$[21.3]$ &
$1.0$               &
$-1.0$              &
$0.5$               &
$-7.0$              &
$1<x<9$             &
carbon              &
(CO)                &
data                &
galactic            &
2,10                \\
\noalign{\smallskip}
18                  &
$[8.8]$--$[21.3]$   &
$2.5$               &
$-0.1$              &
$0.5$               &
$-7.4$              &
$0.1<x<2$           &
carbon              &
(CO)                &
data                &
galactic            &
2,10                \\
\noalign{\smallskip}
\tableline
\end{tabular}
}
\end{center}
\end{table}
\end{landscape}

\citet*{SchroederEtal2003} compute hydrodynamical models in order to generate
synthetic mass-loss rates and infrared colours as the model stars evolve. The
models are computed for initial stellar masses limited to 0.63--1.2 M$_\odot$,
and are only valid for the superwind regime of the highest mass-loss rates.
They also predict H--K and L--M colours but these are less sensitive measures
of the mass-loss rate. Essentially the same models were computed by
\citet{WintersEtal2000}, who showed that although the evolution of mass-loss
rate with infrared colour is qualitatively comparable to that observed by
\citet{Lebertre1997}, the results are offset and display a different slope.

\citet{Groenewegen2006} computes the SEDs for a grid of carbon- and
oxygen-rich AGB stars covering a range in mass-loss rate (similar to what was
done for Fig.\ 1), and convolves these SEDs with a suite of filter curves to
generate the expected optical and infrared photometry. If applying the
appropriate scaling relations as discussed before, these magnitudes can be
compared with actual measurements to estimate the mass-loss rate of a star.

\section{Complications with deriving mass-loss rates from the dust}

The description of the dust-driven wind in {\S 2} is a highly idealized one,
but often the observational constrains are lacking that would justify a more
detailed one. It is thus important to be aware of the caveats, of which I list
a few here.

The dust:gas ratio is difficult to measure except for nearby objects from
which CO line emission can be detected (realising that CO is also a minor
constituent). In well-developed dust-driven winds, the dust condensation seems
to be complete in the sense that all condensable material is locked up in the
grains. This yields a dust:gas mass ratio $\psi\sim0.005$ for stars of an
overall solar metallicity, with the expected scaling of Eq.\ (4). The validity
of this notion is confirmed by agreement between the mass-loss rates predicted
from a formula that was derived for a magellanic sample of oxygen-rich AGB
stars and RSGs, and the mass-loss rates measured from the far-infrared
brightness, for a sample of dust-enshrouded galactic AGB stars and RSGs
\citep{VanloonEtal2005}.

These authors also show (in their Fig.\ 11) that the paradigm breaks down in
the case of giants with warm photospheres or that pulsate less vigorously. In
the case of Betelgeuse this is known to be associated with a low dust:gas
ratio suggesting that the dust condensation process has not reached its full
potential. Although this regime holds vital clues to the dust formation
process, it makes it an extremely hazardous affair to predict the dust:gas
ratio and hence the total mass in the wind. This might gain in relative
importance at low overall metallicity, when the carbon:oxygen ratio approaches
unity, or when Hot Bottom Burning in massive oxygen-rich AGB stars depletes
both carbon and oxygen by burning it into the non-refractive element nitrogen.
The exact composition of the dust and the grain size distribution then also
become more uncertain and with it, the optical constants that determine the
opacity.

In dust-poor winds one can expect a large drift velocity between grains and
gas, and Eq.\ (3) would need to be modified to reflect this
\citep*{HabingEtal1994}. The SED shortward of its peak also depends on the
acceleration of the wind, as well as the radial profile of the dust
condensation process. This can be addressed in detailed non-LTE models for the
line profiles of rotational transitions at radio wavelengths
\citep{DecinEtal2006}. The {\sc dusty} code offers the option to use a
hydrodynamic model for the wind, but not a radial stratification of the dust
properties or dust:gas ratio. Stellar radial pulsations drive shocks through
the inner parts of the envelope which cause density enhancements and
variations in the outflow speed, further modifying the SED. Where the
mass-loss rate is derived from the far-infrared emission from the cold dust,
this can be affected by temporal variations in the wind speed and mass-loss
rate on timescales of centuries to as much as $10^5$ yr --- comparable to the
duration of the superwind phase or the interval between thermal pulses that
occur in the nuclear burning layers of AGB stars. Modeling the full SED can,
in principle, recover some of the mass-loss history \citep{Groenewegen1995},
but complications with the interpretation arise if the wind has swept up a
significant amount of interstellar matter \citep*{VillaverEtal2004}.

The dust envelope is usually assumed to be spherically symmetric. This may be
a good approximation for most AGB stars, but RSGs are often found to display
deviations from this. Worse still, RSGs are encountered near their birth
sites, and their envelopes may be emerged in an anisotropic external
irradiation field \citep*{SchusterEtal2006}. Binarity is sometimes invoked to
explain axi-symmetric outflows from post-AGB objects \citep[e.g., the Red
Rectangle:][]{WatersEtal1998}, but less is known about the effects whilst
still on the AGB. Although the geometry of axi-symmetric envelopes can be
determined from the SED \citep[e.g.,][]{WhitneyEtal2003}, the problem becomes
degenerate if the geometry is more complex but no spatial information is
available --- which is the norm in extragalactic objects
\citep[e.g.,][]{VanloonEtal1999}. This may be especially important for
low-mass red giants, which have a large peculiar velocity with respect to the
ambient interstellar medium and are thus expected to develop bowshocks
\citep*{VillaverEtal2004}.

\section{Mass-loss rates from atomic and molecular infrared transitions}

Molecular hydrogen has transitions at infrared wavelengths, notably at
$\lambda=28.22$ $\mu$m (J=2$\rightarrow$0) with an excitation temperature of
$T_{\rm exc}=510$ K and at $\lambda=17.03$ $\mu$m (J=3$\rightarrow$1) with
$T_{\rm exc}=1015$ K \citep*{BurtonEtal1992}. Unfortunately, the decay times
of these quadrupole transitions are in excess of a century and hence the
expected line emission from circumstellar envelopes is extremely feeble.
Higher-level transitions around $\lambda=2$ $\mu$m have been used to study the
pulsating atmospheres of cool giants \citep{HinkleEtal2000}, but until the
pulsation and its connection with the mass loss are better understood it is
not possible to reliably measure the mass-loss rate from these lines.

Absorption in the Q-branch of the acetylene (C$_2$H$_2$) band at
$\lambda=13.7$ $\mu$m has been used to estimate the mass-loss rates from
magellanic carbon stars \citep{MatsuuraEtal2006}, in agreement with the
mass-loss rates derived from modeling of the SEDs \citep{VanloonEtal2006}. In
its current form it is simply a measurement of the equivalent width and hence
it suffers from dependencies on luminosity, velocity (structure) and chemistry
similar to the optical depth of the dust envelope. But more sophisticated
modeling of the absorption band shapes and line shifts in high-resolution
spectra of the acetylene bands at $\lambda=3$--14 $\mu$m holds great promise
of improvement of the method.

Interesting in particular for dust-poor environments, fine-structure lines of
atomic oxygen at $\lambda=63$ $\mu$m and atomic and singly-ionized iron,
silicon and sulphur at $\lambda=24$--36 $\mu$m can be used to estimate the gas
mass in the line-forming region \citep*{HaasEtal1995,AokiEtal1998}. But, as
with the high-excitation H$_2$ lines, a theory for the mass loss resulting
from such an atmosphere is required in order to infer the mass-loss rate.

An electronic transition, the triplet of atomic helium at $\lambda=1.083$
$\mu$m can be used to trace the bulk flow in the chromosphere that exists
around warm giants ($T_\star>4600$ K), much more effectively than the more
commonly used optical lines of H$\alpha$ and Ca\,{\sc ii} H+K
\citep*{SmithEtal2004}. Unfortunately, the analysis depends sensitively on the
ill-defined structure of the chromosphere.

\section{Concluding remarks}

Several methods are available to determine the mass-loss rate of cool giants
from infrared data. These can be divided in methods based on the dust
extinction and emission, and methods based on the absorption and emission by
molecules and atoms. The ``dust'' methods are ultimately linked to the shape
of the spectral energy distribution, be it through modeling or the use of
derived relationships between the mass-loss rate and far-infrared flux
density, infrared colour or optical extinction. The accuracies obtained are
almost never better than a factor two even in nearby objects with exquisite
data, and can easily be worse than an order of magnitude in distant objects.
More importantly, systematic uncertainties of similar magnitudes arise from
lack of theory, in particular with regard to the structure of the molecular
atmosphere and the dust condensation process.

\acknowledgements 
I would like to warmly thank the organisers and all participants for a very
interesting and pleasant conference in the heart of Europe.


\begin{thebibliography}{}
\bibitem[Aoki et al.(1998)Aoki, Tsuji, \& Ohnaka]{AokiEtal1998}
Aoki, W., Tsuji, T., \& Ohnaka, K. 1998, A\&A, 333, L19
\bibitem[Bjorkman \& Wood(2001)]{BjorkmanWood2001}
Bjorkman, J. E., \& Wood, K. 2001, ApJ, 554, 615
\bibitem[Bouwman(2001)]{Bouwman2001}
Bouwman, J. 2001, PhD thesis, University of Amsterdam
\bibitem[Boyer et al.(2006)]{BoyerEtal2006}
Boyer, M. L., Woodward, C. E., van Loon, J. Th., Gordon, K. D., Evans, A.,
Gehrz, R. D., Helton, L. A., \& Polomski, E. F. 2006, AJ, 132, 1415
\bibitem[Burton et al.(1992)Burton, Hollenbach, \& Tielens]{BurtonEtal1992}
Burton, M. G., Hollenbach, D. J., \& Tielens, A. G. G. M. 1992, ApJ, 399, 563
\bibitem[Decin et al.(2006)]{DecinEtal2006}
Decin, L., Hony, S., de Koter, A., Justtanont, K., Tielens, A. G. G. M., \&
Waters, L. B. F. M. 2006, A\&A, 456, 549
\bibitem[Deutsch(1956)]{Deutsch1956}
Deutsch, A. J. 1956, ApJ, 123, 210
\bibitem[Elitzur et al.(1976)Elitzur, Goldreich, \& Scoville]{ElitzurEtal1976}
Elitzur, M., Goldreich, P., \& Scoville, N. 1976, ApJ, 205, 384
\bibitem[Ercolano et al.(2005)Ercolano, Barlow, \& Storey]{ErcolanoEtal2005}
Ercolano, B., Barlow, M. J., \& Storey, P. J. 2005, MNRAS, 362, 1038
\bibitem[Gehrz \& Woolf(1971)]{GehrzWoolf1971}
Gehrz, R. D., \& Woolf, N. J. 1971, ApJ, 165, 285
\bibitem[Groenewegen(1993)]{Groenewegen1993}
Groenewegen, M. A. T. 1993, PhD thesis, University of Amsterdam
\bibitem[Groenewegen(1995)]{Groenewegen1995}
Groenewegen, M. A. T. 1995, A\&A, 293, 463
\bibitem[Groenewegen(2006)]{Groenewegen2006}
Groenewegen, M. A. T. 2006, A\&A, 448, 181
\bibitem[Groenewegen et al.(2007)]{GroenewegenEtal2007}
Groenewegen, M. A. T., Wood, P. R., Sloan, G. C., Blommaert, J. A. D. L.,
Cioni, M.-R. L., Feast, M. W., Hony, S., Matsuura, M., Menzies, J. W.,
Olivier, E. A., Vanhollebeke, E., van Loon, J. Th., Whitelock, P. A.,
Zijlstra, A. A., Habing, H. J., Lagadec, E. 2007, submitted to MNRAS
\bibitem[Guandalini et al.(2006)]{GuandaliniEtal2006}
Guandalini, R., Busso, M., Ciprini, S., Silvestro, G., \& Persi, P. 2006,
A\&A, 445, 1069
\bibitem[Haas et al.(1995)Haas, Glassgold, \& Tielens]{HaasEtal1995}
Haas, M. R., Glassgold, A. E., \& Tielens, A. G. G. M. 1995, in Airborne
Astronomy Symposium on the Galactic Ecosystem: From Gas to Stars to Dust. ASP
Conference Series, Vol.\ 73, p.\ 397
\bibitem[Habing et al.(1994)Habing, Tignon, \& Tielens]{HabingEtal1994}
Habing, H. J., Tignon, J., \& Tielens, A. G. G. M. 1994, A\&A, 286, 523
\bibitem[Hinkle et al.(2000)]{HinkleEtal2000}
Hinkle, K. H., Aringer, B., Lebzelter, T., Martin, C. L., \& Ridgway, S. T.
2000, A\&A, 363, 1065
\bibitem[Ivezi\'c \& Elitzur(1995)]{IvezicElitzur1995}
Ivezi\'c, \v{Z}., \& Elitzur, M. 1995, ApJ, 445, 415
\bibitem[Ivezi\'c et al.(1999)Ivezi\'c, Elitzur, \& Nenkova]{IvezicEtal1999}
Ivezi\'c, \v{Z}, Nenkova, M., \& Elitzur, M. 1999, User manual for {\sc
dusty}. University of Kentucky Internal Report
\bibitem[Jackson et al.(2007)]{JacksonEtal2007}
Jackson, D. C., Skillman, E. D., Gehrz, R. D., Polomski, E., \& Woodward, C.
E. 2007, ApJ, in press
\bibitem[Jura(1986)]{Jura1986}
Jura, M. 1986, IrAJ, 17, 322
\bibitem[Jura(1987)]{Jura1987}
Jura, M. 1987, ApJ, 313, 743
\bibitem[Knapp \& Morris(1985)]{KnappMorris1985}
Knapp, G. R., \& Morris, M. 1985, ApJ, 292, 640
\bibitem[Le Bertre(1997)]{Lebertre1997}
Le Bertre, T. 1997, A\&A, 324, 1059
\bibitem[Le Bertre \& Winters(1998)]{LebertreWinters1998}
Le Bertre, T., \& Winters, J. M. 1998, A\&A, 334, 173
\bibitem[Le Sidaner \& Le Bertre(1996)]{LesidanerLebertre1996}
Le Sidaner, P., \& Le Bertre, T. 1996, A\&A, 314, 896
\bibitem[Leung(1975)]{Leung1975}
Leung, C. M. 1975, ApJ, 199, 340
\bibitem[Marshall et al.(2004)]{MarshallEtal2004}
Marshall, J. R., van Loon, J. Th., Matsuura, M., Wood, P. R., Zijlstra, A. A.,
\& Whitelock, P. A. 2004, MNRAS, 355, 1348
\bibitem[Matsuura et al.(2006)]{MatsuuraEtal2006}
Matsuura, M., Wood, P. R., Sloan, G. C., Zijlstra, A. A., van Loon, J. Th.,
Groenewegen, M. A. T., Blommaert, J. A. D. L., Cioni, M.-R. L., Feast, M. W.,
Habing, H. J., Hony, S., Lagadec, E., Loup, C., Menzies, J. W., Waters, L. B.
F. M., \& Whitelock, P. A. 2006, MNRAS, 371, 415
\bibitem[Olnon et al.(1984)]{OlnonEtal1984}
Olnon, F. M., Habing, H. J., Baud, B., Pottasch, S. R., de Jong, T., \&
Harris, S. 1984, ApJ, 278, L41
\bibitem[Reimers(1975)]{Reimers1975}
Reimers, D. 1975, in $19^{\rm th}$ International Astrophysics Colloquium.
Societ\'{e} Royale des Sciences de Li\`{e}ge, Memoires Vol.\ 8, p.\ 369
\bibitem[Rowan-Robinson(1980)]{Rowanrobinson1980}
Rowan-Robinson, M. 1980, ApJS, 44, 403
\bibitem[Schr\"oder et al.(2003)Schr\"oder, Wachter, \&
         Winters]{SchroederEtal2003}
Schr\"oder, K.-P., Wachter, A., \& Winters, J. M. 2003, A\&A, 398, 229
\bibitem[Schuster et al.(2006)Schuster, Humphreys, \&
         Marengo]{SchusterEtal2006}
Schuster, M. T., Humphreys, R. M., \& Marengo, M. 2006, AJ, 131, 603
\bibitem[Schutte \& Tielens(1989)]{SchutteTielens1989}
Schutte, W. A., \& Tielens, A. G. G. M. 1989, ApJ, 343, 369
\bibitem[Smith et al.(2004)Smith, Dupree, \& Strader]{SmithEtal2004}
Smith, G. H., Dupree, A. K., \& Strader, J. 2004, PASP, 116, 819
\bibitem[van Loon(2006)]{Vanloon2006}
van Loon, J. Th. 2006, in Stellar Evolution at Low Metallicity: Mass Loss,
Explosions, Cosmology. ASP Conference Series, Vol.\ 353, p.\ 211
\bibitem[van Loon et al.(1999)]{VanloonEtal1999}
van Loon, J. Th., Groenewegen, M. A. T., de Koter, A., Trams, N. R., Waters,
L. B. F. M., Zijlstra, A. A., Whitelock, P. A., \& Loup, C. 1999, A\&A, 351,
559
\bibitem[van Loon et al.(2005)]{VanloonEtal2005}
van Loon, J. Th., Cioni, M.-R. L., Zijlstra, A. A., \& Loup, C. 2005, A\&A,
438, 273
\bibitem[van Loon et al.(2006)]{VanloonEtal2006}
van Loon, J. Th., Marshall, J. R., Cohen, M., Matsuura, M., Wood, P. R.,
Yamamura, I., \& Zijlstra, A. A. 2006, A\&A, 447, 971
\bibitem[Villaver et al.(2004)Villaver, Garc\'{\i}a-Segura, \&
         Manchado]{VillaverEtal2004}
Villaver, E., Garc\'{\i}a-Segura, G., \& Manchado, A. 2004, RevMexA\&A, 22,
140
\bibitem[Waters et al.(1998)]{WatersEtal1998}
Waters, L. B. F. M., Waelkens, C., Van Winckel, H., Molster, F. J., Tielens,
A. G. G. M., van Loon, J. Th., Morris, P. W., Cami, J., Bouwman, J., de Koter,
A., de Jong, T., \& de Graauw, Th. 1998, Nature, 391, 868
\bibitem[Whitelock et al.(1991)Whitelock, Feast, \&
         Catchpole]{WhitelockEtal1991}
Whitelock, P., Feast, M., \& Catchpole, R. 1991, MNRAS, 248, 276
\bibitem[Whitelock et al.(1994)]{WhitelockEtal1994}
Whitelock, P. A., Menzies, J., Feast, M., Marang, F., Carter, B., Roberts, G.,
Catchpole, R., \& Chapman, J. 1994, MNRAS, 267, 711
\bibitem[Whitney et al.(2003)]{WhitneyEtal2003}
Whitney, B. A., Wood, K., Bjorkman, J. E., \& Wolff, M. J. 2003, ApJ, 591,
1049
\bibitem[Winters et al.(2000)]{WintersEtal2000}
Winters, J. M., Le Bertre, T., Jeong, K. S., Helling, Ch., \& Sedlmayr, E.
2000, A\&A, 361, 641
\bibitem[Wood et al.(1992)]{WoodEtal1992}
Wood, P. R., Whiteoak, J. B., Hughes, S. M. G., Bessell, M. S., Gardner, F.
F., \& Hyland, A. R. 1992, ApJ, 397, 552
\bibitem[Yorke(1980)]{Yorke1980}
Yorke, H. W. 1980, A\&A, 86, 286
\bibitem[Zijlstra et al.(1996)]{ZijlstraEtal1996}
Zijlstra, A. A., Loup, C., Waters, L. B. F. M., Whitelock, P. A., van Loon, J.
Th., \& Guglielmo, F. 1996, MNRAS, 279, 32
\end{thebibliography}
\end{document}